\documentclass[aps,prl,preprint,superscriptaddress]{revtex4}
\usepackage{graphicx}
\usepackage{bm}
\usepackage{color}
\usepackage{amsfonts}
\usepackage{isomath}
\usepackage{amsmath}
\usepackage{amsthm}
\usepackage{amsfonts}
\usepackage{amssymb}
\usepackage{fdsymbol}
\usepackage{braket}
\usepackage{gensymb}

\def\VSe2{1$T$-VSe$_2$}
\def\TaS2{1$T$-TaS$_2$}
\def\TiSe2{1$T$-TiSe$_2$}

\def\V{V~3\textit{d} }
\def\Se{Se~4\textit{p} }

\begin{document}

\title{Hot carrier-assisted switching of the electron-phonon interaction in~\VSe2}
\author{Paulina Majchrzak}
\affiliation{Department of Physics and Astronomy, Interdisciplinary Nanoscience Center, Aarhus University,
8000 Aarhus C, Denmark}
\affiliation{Central Laser Facility, STFC Rutherford Appleton Laboratory, Harwell 0X11 0QX, United Kingdom}
\author{Sahar Pakdel}
\affiliation{Department of Physics and Astronomy, Interdisciplinary Nanoscience Center, Aarhus University,
8000 Aarhus C, Denmark}
\author{Deepnarayan Biswas}
\affiliation{Department of Physics and Astronomy, Interdisciplinary Nanoscience Center, Aarhus University,
8000 Aarhus C, Denmark}
\author{Alfred J. H. Jones}
\affiliation{Department of Physics and Astronomy, Interdisciplinary Nanoscience Center, Aarhus University,
8000 Aarhus C, Denmark}
\author{Klara Volckaert}
\affiliation{Department of Physics and Astronomy, Interdisciplinary Nanoscience Center, Aarhus University,
8000 Aarhus C, Denmark}
\author{Igor Markovi\'c}
\affiliation{SUPA, School of Physics and Astronomy, University of St Andrews, St Andrews KY16 9SS, United Kingdom}
\affiliation{Max Planck Institute for Chemical Physics of Solids, N\"othnitzer Stra{\ss}e 40, 01187 Dresden, Germany}
\author{Federico Andreatta}
\affiliation{Department of Physics and Astronomy, Interdisciplinary Nanoscience Center, Aarhus University,
8000 Aarhus C, Denmark}
\author{Raman Sankar}
\affiliation{Institute of
Physics, Academia Sinica, Taipei 11529, Taiwan}
\author{Chris Jozwiak}
\affiliation{Advanced Light Source, E. O. Lawrence Berkeley National Laboratory, Berkeley, California 94720, USA}
\author{Eli Rotenberg}
\affiliation{Advanced Light Source, E. O. Lawrence Berkeley National Laboratory, Berkeley, California 94720, USA}
\author{Aaron Bostwick}
\affiliation{Advanced Light Source, E. O. Lawrence Berkeley National Laboratory, Berkeley, California 94720, USA}
\author{Charlotte E. Sanders}
\affiliation{Central Laser Facility, STFC Rutherford Appleton Laboratory, Harwell 0X11 0QX, United Kingdom}
\author{Yu Zhang}
\affiliation{Central Laser Facility, STFC Rutherford Appleton Laboratory, Harwell 0X11 0QX, United Kingdom}
\author{Gabriel Karras}
\affiliation{Central Laser Facility, STFC Rutherford Appleton Laboratory, Harwell 0X11 0QX, United Kingdom}
\author{Richard T. Chapman}
\affiliation{Central Laser Facility, STFC Rutherford Appleton Laboratory, Harwell 0X11 0QX, United Kingdom}
\author{Adam Wyatt}
\affiliation{Central Laser Facility, STFC Rutherford Appleton Laboratory, Harwell 0X11 0QX, United Kingdom}
\author{Emma Springate}
\affiliation{Central Laser Facility, STFC Rutherford Appleton Laboratory, Harwell 0X11 0QX, United Kingdom}
\author{Jill A. Miwa}
\affiliation{Department of Physics and Astronomy, Interdisciplinary Nanoscience Center, Aarhus University,
8000 Aarhus C, Denmark}
\author{Philip Hofmann}
\affiliation{Department of Physics and Astronomy, Interdisciplinary Nanoscience Center, Aarhus University,
8000 Aarhus C, Denmark}
\author{Phil D. C. King}
\affiliation{SUPA, School of Physics and Astronomy, University of St Andrews, St Andrews KY16 9SS, United Kingdom}
\author{Nicola Lanata}
\affiliation{Department of Physics and Astronomy, Interdisciplinary Nanoscience Center, Aarhus University,
8000 Aarhus C, Denmark}
\author{Young Jun Chang}
\affiliation{Department of Physics, University of Seoul, Seoul 02504, Republic of Korea}
\author{S{\o}ren~Ulstrup}
\email[Correspondence to S{\o}ren~Ulstrup: ]{ulstrup@phys.au.dk}  
\affiliation{Department of Physics and Astronomy, Interdisciplinary Nanoscience Center, Aarhus University,
8000 Aarhus C, Denmark}

\begin{abstract}
We apply an intense infrared laser pulse in order to perturb the electronic and vibrational states in the three-dimensional charge density wave material \VSe2. Ultrafast snapshots of the light-induced hot carrier dynamics and non-equilibrium quasiparticle spectral function are collected using time- and angle-resolved photoemission spectroscopy. The hot carrier temperature and time-dependent electronic self-energy are extracted from the time-dependent spectral function, revealing that incoherent electron-phonon interactions heat the lattice above the charge density wave critical temperature on a timescale of $(200 \pm 40)$~fs. Density functional perturbation theory calculations establish that the presence of hot carriers alters the overall phonon dispersion and quenches efficient low-energy acoustic phonon scattering channels, which results in a new quasi-equilibrium state that is experimentally observed.
\end{abstract}

\maketitle

The metallic transition metal dichalcogenide (TMDC) \VSe2 exhibits a charge density wave (CDW) transition around a temperature below $T_c =110$~K with  in-plane commensurate (4 $\times$ 4) and out-of-plane incommensurate components \cite{VanBruggen:1976,Tsutsumi:1982,Eagleshamt:1986}. Due to its single-band Fermi surface derived from vanadium 3$d$ states, this CDW transition is commonly viewed as a three-dimensional (3D) analogue to the one-dimensional Peierls insulator \cite{Terashima:2003, Sato:2004, Strocov:2012}. However, a weak lattice distortion and poor nesting condition lead to very subtle spectroscopic signatures of the CDW state \cite{Henke:2019}. A narrow partial gap of 24 meV has been detected by tunneling spectroscopy \cite{Jolie:2019}, while no replica bands around the CDW wave vector have been observed by angle-resolved photoemission (ARPES) measurements \cite{Terashima:2003, Sato:2004, Strocov:2012}.

The CDW transition temperature and electronic gap of \VSe2 can be significantly enhanced by reducing the thickness of the material to a single layer \cite{Pasztor:2017,Duvjir:2018, Feng:2018, Chen:2018} or by applying a high pressure to the bulk \cite{Feng:2020}, eventually inducing a superconducting phase~\cite{Sahoo:2020}. An unexplored alternative strategy of modifying the CDW phase relies upon optically exciting the material with an intense laser pulse and measuring the time-dependent evolution of the electronic and lattice degrees of freedom underpinning the CDW phase. It is thereby possible to disentangle  carrier-carrier screening, electron-optical phonon coupling and scattering involving low-energy soft acoustic phonons \cite{Zhang:2017, Si:2020} in the excitation and ensuing relaxation processes and to determine their role on dynamically driven phase transitions.

To exploit this strategy, we employ time-resolved ARPES (TR-ARPES) where an infrared pump pulse excites the electrons around the Fermi level, $E_F$, followed by a femtosecond extreme ultraviolet (XUV) probe pulse after a variable time delay, leading to photoemission from the excited state. By tracking the subsequent relaxation of the excited state quasiparticle spectrum, it is possible to extract the hot carrier temperature and to determine the temporal evolution of the electron-phonon interaction \cite{Perfetti:2007,Johannsen:2013,Perfetti:2006,Rohwer:2011,Petersen:2011,Hellmann:2012,Monney:2016}. Our single crystals of \VSe2 have been grown using the chemical vapor transport method with I$_2$ as a transport agent \cite{Sayers:2020, Feroze:2020}. Unless otherwise stated, all photoemission experiments have been carried out with the sample held at a temperature of $\approx$70~K, ensuring that at equilibrium, the material is measured in its CDW phase.

\begin{figure} [t!]
	\includegraphics[width=0.79\textwidth]{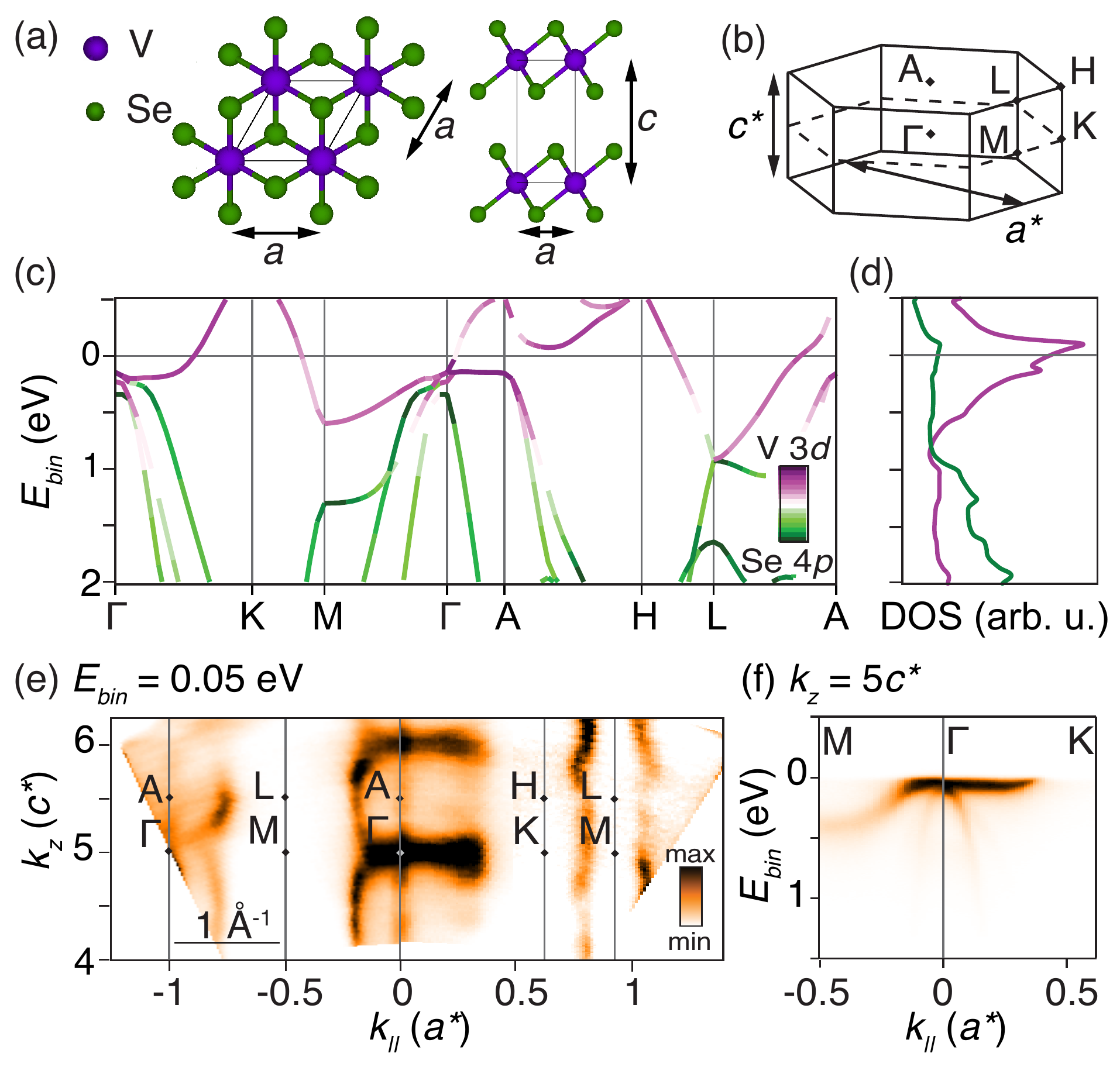}\\
\caption{ 
	(a) Crystal structure of \VSe2 with lattice parameters $a$ and $c$. (b) Hexagonal BZ and high symmetry points with in-plane repetition $a^{\ast} = 4\pi/\sqrt{3}a$ and out-of-plane repetition $c^{\ast} = 2\pi/c$. Double-headed arrows illustrate dimensions.
	(c) Band structure projected onto \V and Se~4$p$ orbitals calculated by DFT. 
	(d) Corresponding density of states (DOS) separated in contributions from the \V (purple curve) and Se~4$p$ (green curve) orbitals.
	(e) Photoemission intensity for the {$(k_{//},k_z)$}-plane spanning the given high symmetry points at a binding energy of 0.05 eV.
	(f) ARPES cut obtained for $k_z = 5c^*$, corresponding to the $\mathrm{M-\Gamma-K}$ direction.}
	\label{fig1}
\end{figure}

In order to select a 2D ($E_{bin},k_{//}$)-cut of the photoemission intensity that tracks excited carriers in the dispersion around $E_F$, we first characterize the static photoemission properties of \VSe2. In the normal state, the system adopts the tetragonal crystal structure in Fig. \ref{fig1}(a) with the corresponding hexagonal Brillouin zone (BZ) shown in Fig. \ref{fig1}(b). The electronic band structure and density of states (DOS) of this structure are presented in Figs. \ref{fig1}(c)-(d) \cite{SMAT,Blaha:2018,Perdew:1996}, revealing several $E_F$ crossings of the bands with a dominant contribution from V 3$d$ orbitals. This normal state electronic structure is compared with our static ARPES measurements of the \VSe2 dispersion in the CDW phase performed with synchrotron radiation over a photon energy range of 65-160 eV, as seen in Figs. \ref{fig1}(e)-(f). In the  {$(k_{//},k_z)$}-dependent photoemission intensity around high symmetry points of the bulk BZ near $E_F$ in Fig. \ref{fig1}(e), a strikingly high signal appears around $\mathrm{\Gamma}$ \cite{SMAT}. This is seen in Fig. \ref{fig1}(f) to emerge from a flat part of the dispersion that crosses $E_F$ towards $\mathrm{K}$. This closely resembles the corresponding DFT dispersion of the V 3$d$ band in the normal state in Fig. \ref{fig1}(c). Monitoring the response of the system to an optical excitation in this direction thus provides access to excited carriers following the band and thereby the broadening of the Fermi-Dirac (FD) function caused by an elevated electronic temperature, $T_e$, resulting from thermalization of the carriers \cite{Ulstrup:2014}. We do not find any clear manifestation of the CDW in the static ARPES intensity \cite{Strocov:2012,SMAT}. 

\begin{figure} [t!]
	\includegraphics[width=0.79\textwidth]{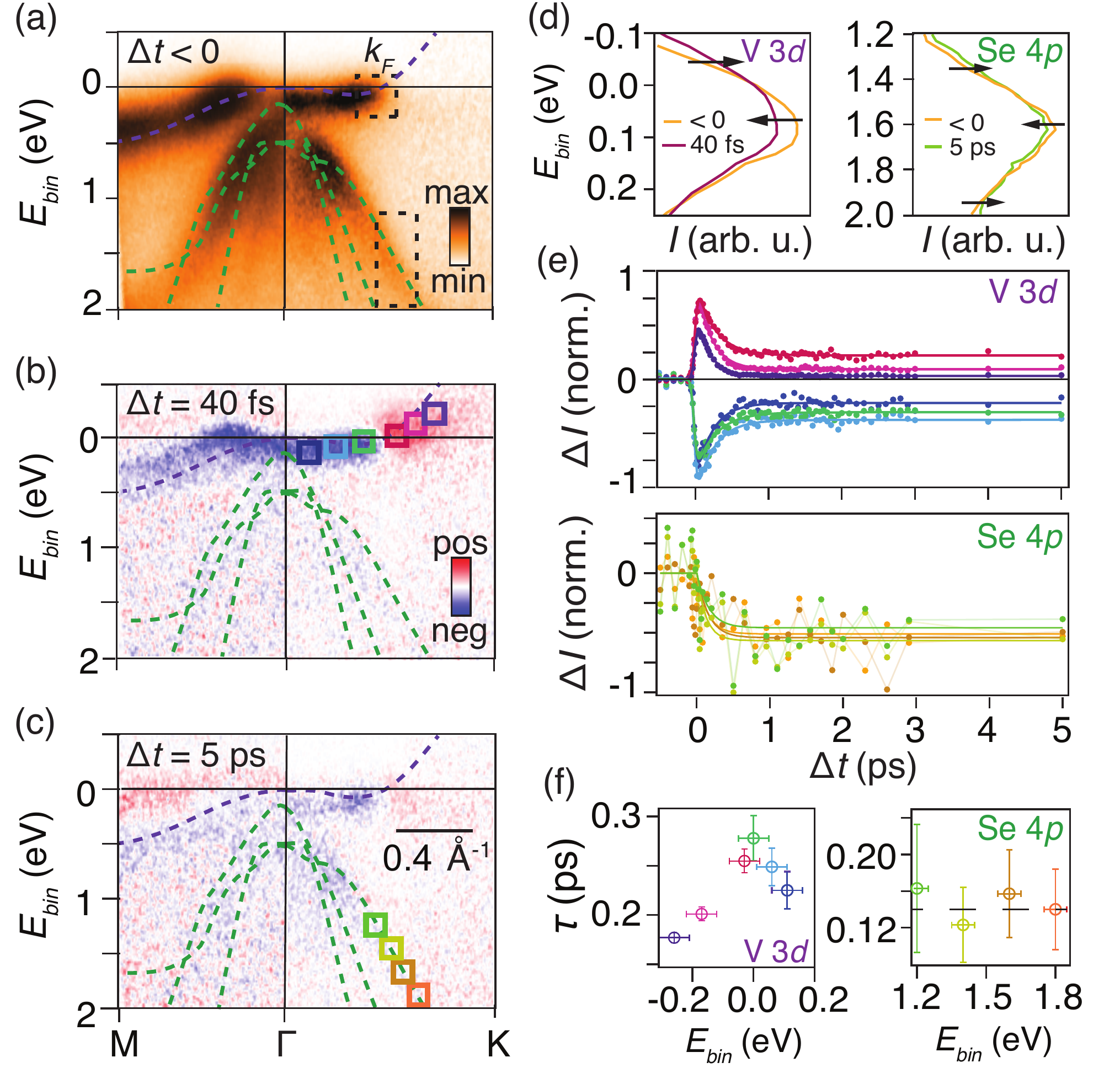}\\
	\caption{
	 (a) TR-ARPES intensity along $\mathrm{M-\Gamma-K}$ before optical excitation ($\Delta t < 0$). 
	 (b)-(c) Intensity difference between a spectrum obtained at the given time delay and the equilibrium spectrum in (a). The dashed lines are DFT bands, which encode V 3$d$ (purple lines) and Se~4$p$ (green lines) character.
	 (d) EDCs integrated over the momentum range indicated in (a) by dashed boxes around the Fermi energy (left panel) and around the Se~4$p$ bands (right panel). The arrows indicate the redistribution of intensity following photoexcitation. 
	 (e) Intensity difference integrated over the $(E_{bin},k_{//})$-regions demarcated by correspondingly colored boxes in (b)-(c). The smooth curves are fits to an exponential.
	 (f) Time constants for the exponential fits in (e). The markers have been colored according to the corresponding boxes in (b)-(c) and curves in (e). The dashed line in the right panel represents an average time constant.}
	\label{fig2}
\end{figure}

We now focus on TR-ARPES measurements performed along the selected $k_{//}$-cut using an XUV pulse with a photon energy of 29.6 eV obtained by high harmonic generation, as demonstrated in Fig. \ref{fig2}(a). The features are well-described by the overlaid DFT bands along $\mathrm{M-\Gamma-K}$ and closely resemble the synchrotron data in Fig. \ref{fig1}(f) with differences in intensity arising from the change of photon energy and the different photoemission geometries \cite{SMAT}. Figures \ref{fig2}(b)-(c) present the intensity difference between an excited state spectrum at the given time delays, $\Delta t$, and the equilibrium spectrum in Fig. \ref{fig2}(a) following optical excitation with a 1.55 eV pump pulse with a repetition rate of 1~kHz and a fluence of 2.2 mJ/cm$^{2}$. The latter is chosen to achieve a clear signal above the noise for the excited state while avoiding space-charge effects in the spectra. The time resolution is 40 fs.

The red (blue) regions in Figs. \ref{fig2}(b)-(c) correspond to photoemission intensity gain (loss). Immediately after the excitation, at $\Delta t = 40$ fs, the signal along $\mathrm{\Gamma-K}$ exhibits a loss below $E_F$ that is compensated by a gain in the band above $E_F$, indicating the presence of excited holes and electrons. In contrast, along $\mathrm{\Gamma-M}$, where the bands do not cross $E_F$, we merely observe a loss signal below $E_F$. At $\Delta t = 5$ ps, a significant intensity difference is still observable around the $E_F$ crossing. As most clearly seen by the momentum-integrated energy distribution curves (EDCs) in Fig. \ref{fig2}(d), an additional loss (gain) of intensity is seen around the center (tail) of all the bands, indicating an overall broadening effect. The left panel presents EDCs around the Fermi wave vector, $k_F$, in the equilibrium ($t<$ 0) and at the peak of excitation (40 fs), demonstrating the filling (depletion) of electrons above (below) $E_F$. The right panel shows EDCs for the Se~4$p$ bands towards higher binding energy at a long delay (5 ps), exhibiting a significant broadening effect, similar to the corresponding intensity difference in Fig. \ref{fig2}(c). We emphasize that we do not observe any rigid energy shifts in the EDCs, thereby ruling out a significant influence from extrinsic space-charge effects. 

The detailed time-dependence of these signals is investigated by integrating the intensity difference within the $(E_{bin},k_{//})$-regions following the bands, as marked by colored boxes, in Figs.  \ref{fig2}(b)-(c). The resulting $\Delta I(t)$ curves are shown in Fig. \ref{fig2}(e). The upper panel presents the intensity changes in the V 3$d$ states around $E_F$. A sharp transient, nearly symmetric for holes and electrons, is observed, followed by a brief relaxation period before the signal reaches a metastable situation that remains far from the equilibrium signal on the timescale we probe. In the lower panel we inspect the time-dependence of the broadening of the Se~4$p$ states. Interestingly, this effect grows after the excitation and settles at a fixed level without any sign of recovery for the probed time delays. Exponential function fits of the relaxation part in the V~3$d$ states and the change of intensity in the Se~4$p$ states provide the energy-dependence of the time constants shown in Fig. \ref{fig2}(f). In the V~3$d$ states the relaxation slows down towards the Fermi energy, reflecting the behavior expected for hot carriers that follow the FD distribution \cite{Crepaldi:2012,Johannsen:2013}. The states towards higher binding energies broaden on a time scale of around 150 fs, which is independent of the selected ($E_{bin},k_{//}$)-window as seen in the right panel in Fig. \ref{fig2}(f).  

\begin{figure} [t!]
	\includegraphics[width=0.70\textwidth]{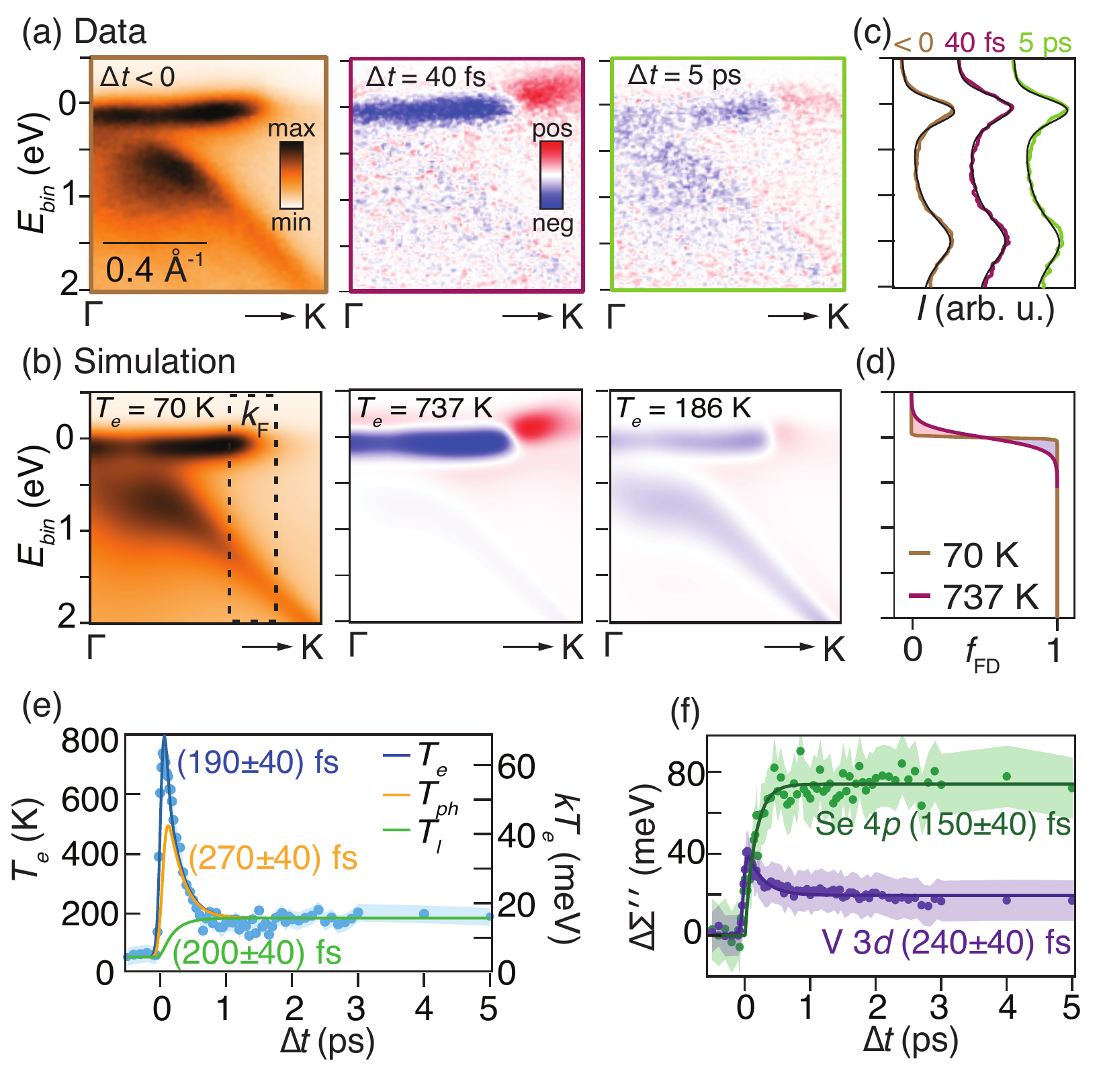}\\
	\caption{
	(a) ARPES intensity in equilibrium (left panel) and intensity difference at $\Delta t$ = 40 fs (middle panel) and $\Delta t$ = 5~ps (right panel) along  $\mathrm{\Gamma-K}$. 
	(b) Simulations of the ARPES intensity and the resulting intensity difference and electronic temperature, $T_e$, from fits of the data displayed in the same column in (a).
	(c) Measured (colored curves) and simulated (smooth black curves) EDCs integrated over the momentum range indicated by the black dashed box in the left panel in (b) for the given time delays.
	(d) Fermi-Dirac distributions at the extreme electronic temperatures.
    (e) Temporal evolution of $T_e$ and the temperature of strongly coupled modes, $T_{ph}$, and remaining lattice, $T_l$, determined via a three-temperature model \cite{SMAT}. Markers correspond to $T_e$ values obtained from the intensity simulations. The time constants are given for the initial relaxations of $T_e$ and $T_{ph}$ and heating of $T_l$.
	(f) Time dependent change of imaginary part of the electronic self-energy for the \V (purple markers) and \Se (green markers) states determined from the spectral function fits. Smooth curves are fits to functions with the given time constants for the exponential components.
	The shaded regions in (e)-(f) indicate the error associated with extracted values.}
	\label{fig3}
\end{figure}

In order to disentangle the contributions of the \VSe2 spectral function and the hot carrier population to the intensity difference described above, we apply the proportional relation between the photoemission intensity, ${\cal I}$, and the product ${\cal A}(E_{bin},k_{//})f_{FD}(E,T_e)$, where $T_e$ enters via the FD distribution, $f_{FD}$. The spectral function is described by ${\cal A}(E_{bin},{k_{//}}) = \pi^{-1} \Sigma^{\prime\prime}/([E_{bin}-\epsilon(k_{//}) - {\Sigma^{\prime}} ]^2 +{\Sigma^{\prime\prime}}^2)$, where $\epsilon(k_{//})$ is the bare dispersion, $\Sigma^{\prime}$ is the real and $\Sigma^{\prime\prime}$ the imaginary part of the electronic self-energy \cite{Nechaev:2009,Ulstrup:2014, Andreatta:2019,Biswas:2020}. In order to simulate the measured 2D image of ${\cal I}(E_{bin},k_{//})$, we use the DFT bands overlaid in Fig. \ref{fig2}(a)-(c) in place of the bare dispersion $\epsilon(k_{//})$. Since no time-dependent shifts in the band positions are observed in the data, we approximate $\Sigma^{\prime}$ as a time-independent constant. The time-dependent broadening effects are encoded in $\Sigma^{\prime\prime}$, which is taken as independent of energy and momentum within the V 3$d$ and Se 4$p$ states, justified by the analysis discussed in connection with Figs. \ref{fig2}(e)-(f). By fitting $T_e$ and $\Sigma^{\prime\prime}$ in the simulated intensity to the measured ARPES intensity at each time delay, we are able to quantitatively extract the temporal evolution of the hot carrier dynamics and the quasiparticle scattering rate described by these two parameters \cite{Andreatta:2019,Biswas:2020,SMAT}. 

The simulated intensity within this model provides an excellent description of the data, as seen by comparing Figs. \ref{fig3}(a)-(b), as well as the example EDCs in Fig. \ref{fig3}(c). Since the data are well-described by a fit including a FD function with an elevated temperature at all time delays, we can assume that the carriers thermalize within the 40 fs timescale we can resolve. A maximum electronic temperature of $(737 \pm 30)$~K is observed immediately after excitation, leading to a significant redistribution of carriers in the V 3$d$ states, as seen via the change of width of the FD function in Fig. \ref{fig3}(d).

Figure \ref{fig3}(e) presents the time-dependent change of $T_e$. Following a step-like response to the optical excitation, an initial relaxation on a timescale of $(190 \pm 40)$~fs occurs before a new thermal equilibrium with an elevated $T_e$ is reached. The initial fast relaxation period is explained by incoherent coupling between the hot electrons and phonons. Using a three-temperature model (3TM) incorporating a fraction of strongly coupled Einstein modes and an anharmonic decay of these modes involving the remaining lattice provides an estimate of the temporal evolution of the strongly coupled phonon temperature, $T_{ph}$, and the lattice temperature, $T_l$, as shown in Fig. \ref{fig3}(e)  \cite{Allen:1987, Kampfrath:2005,Perfetti:2007,DalConte:2012,Johannsen:2013,Yadav:2010,SMAT}. The fraction of Einstein modes is centered at an energy of  21~meV, which is explained below. The lattice reaches a temperature of $(182 \pm 30)$~K on a time scale of $(200 \pm 40)$~fs, such that the metastable thermal equilibrium between all three temperatures occurs above $T_c =110$~K. 

The temporal behavior of the self-energy for the V 3$d$ states in Fig. \ref{fig3}(f) resembles closely that of $T_e$, reflecting the changing number of available scattering channels as a result of initial depopulation and subsequent incomplete repopulation of these electronic states. The distinct time-dependent evolution of $\Sigma^{\prime\prime}$ for the Se 4$p$ states at higher binding energies, seen in Fig. \ref{fig3}(f), is more striking since the population of these states remains fixed. Interestingly, the self-energy increases with a very similar time constant as found for the heating of the lattice in Fig. \ref{fig3}(e). The timescale and origin of the increase in $\Sigma^{\prime\prime}$ is therefore explained by the new scattering channels for the decay of the photohole involving absorption of the excited hot phonons. 

\begin{figure} [t!]
	\includegraphics[width=0.79\textwidth]{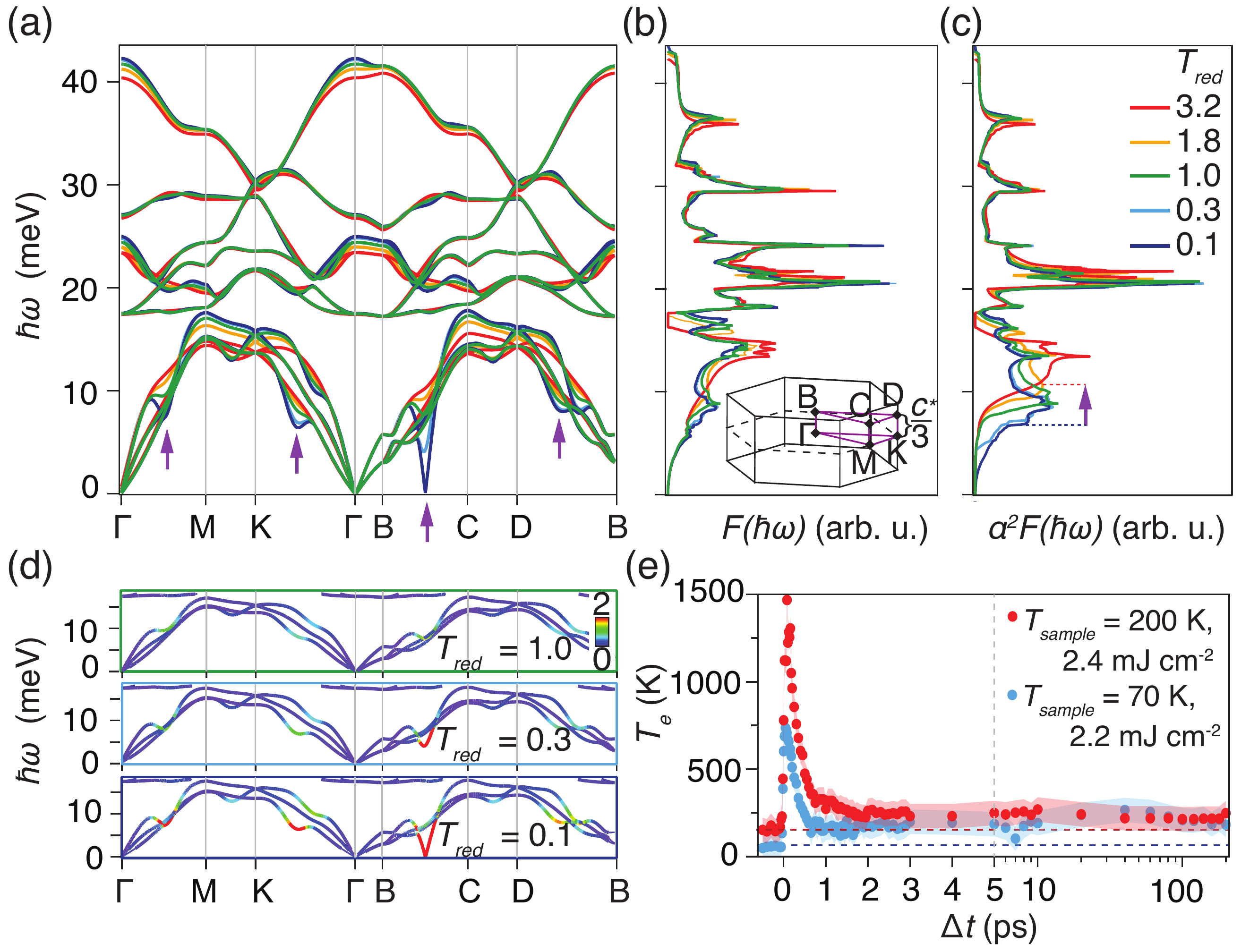}\\
		\caption{
	(a)-(c) Density functional perturbation theory results for (a) phonon dispersion, (b) phonon DOS and (c) Eliashberg function. The inset in (b) displays the hexagonal BZ with the momentum-path in (a) indicated by purple lines. The colors of the curves correspond to the reduced electronic temperatures, $T_{red} = (T_e - T_c)/T_c$, stated in (c). Purple arrows in (a) and (c) indicate phonon energies that depend significantly on $T_{red}$.
	(d) Acoustic phonon dispersion with EPC strength encoded by the color scale at the given values of $T_{red}$.
	(e) Temporal evolution of $T_e$ measured by TR-ARPES with the sample kept at 70 K (blue markers) and 200 K (red markers) at the given pump laser fluence. Note the logarithmic scale after $\Delta t$ = 5 ps. The shaded area indicates the error bar associated with the extraction of $T_e$.
}
	\label{fig4}
\end{figure}

We explore how the electron-phonon interactions are altered through the excitation and relaxation processes by calculating the phonon dispersion and associated mode-resolved electron-phonon coupling (EPC) using density functional perturbation theory \cite{SMAT,migdal1958interaction,eliashberg1960interactions,Giannozzi:2009,Vanderbilt:1990,Grimme:2006,Bayard:1976,Marzari:1999}. The phonon dispersion and associated DOS are shown in Figs. \ref{fig4}(a)-(b) for a range of reduced electronic temperatures, $T_{red} = (T_e - T_c)/T_c$, reflecting the impact of electronic screening on the lattice vibrations \cite{Kohn:1959,Chan:1973,Duong:2015}. The selected path in $\bm{q}$-space is indicated by the BZ sketch in Fig. \ref{fig4}(b). At temperatures close to $T_c$, the phonon dispersion contains a number of soft acoustic modes, seen as cusps in Fig. \ref{fig4}(a). Around the CDW wave vector, the soft mode frequency becomes imaginary below $T_c$, creating an instability towards lattice deformation \cite{SMAT}. The EPC-weighted phonon DOS given by the Eliashberg function, $\alpha^2 F$, reveals a sharp peak from the optical branches at 21~meV in Fig. \ref{fig4}(c). The combination of a high phonon energy and a large EPC strength leads to a strong weight of these optical modes in the energy transfer rate equations that form the basis for the 3TM model discussed in connection with Fig. \ref{fig3}(e) \cite{Allen:1987}. As $T_{red}$ increases, the soft modes disappear, and the low-energy peaks in $\alpha^2 F$ shift to higher energies (see purple arrows in Figs. \ref{fig4}(a) and \ref{fig4}(c)), consistent with the calculated band-resolved phonon dispersion and EPC at selected $T_{red}$ in Fig. \ref{fig4}(d). The phase space of soft modes and the coupling to these modes thus weakens with increasing $T_e$, removing possible efficient cooling channels for the hot carriers.

Finally, we consider the time-dependence of $T_e$ up to 200~ps after the optical excitation for the sample held at temperatures below and above $T_c$, as shown in Fig. \ref{fig4}(e). For the sample above $T_c$ (200~K), the maximum in the electronic temperature is higher than for the sample below $T_c$ (70 K), which we ascribe to a higher laser fluence used in the former case. Overall, however, we observe a very similar time dependent response in the two situations, with a single rapid decay of $T_e$ followed by a metastable state. This precludes an explanation of the metastable state solely involving an energy barrier between the normal state and the CDW \cite{Wall:2012,Diego:2020}. Instead, the optical excitation establishes a new quasi-equilibrium state with a phonon dispersion locked by the elevated $T_e$, which persists for at least hundreds of picoseconds.

Our results demonstrate that a femtosecond optical excitation of \VSe2 leads to a rapidly thermalized hot carrier distribution, which cools on a $(190 \pm 40)$~fs timecale via incoherent electron-phonon interactions that in turn heat the lattice above the CDW transition temperature. The hot carriers dynamically affect the low-energy soft phonon spectrum and EPC strength, leaving the system in a quasi-equilibrium. Such long-lived light-induced states could play a significant role for the dynamics in other CDW materials where the presence of hot carriers leads to the quenching of the soft phonon modes that participate in stabilizing the CDW state.

\begin{acknowledgments}
We thank Phil Rice, Alistair Cox and David Rose for technical support during the Artemis beamtime. We gratefully acknowledge funding from VILLUM FONDEN through the Young Investigator Program (Grant. No. 15375) and the Centre of Excellence for Dirac Materials (Grant. No. 11744), the Danish Council for Independent Research, Natural Sciences under the Sapere Aude program (Grant Nos.  DFF-9064-00057B and DFF-6108-00409) and the Aarhus University Research Foundation. This work is also supported by National Research Foundation (NRF) grants funded by the Korean government (nos. NRF-2020R1A2C200373211 and 2019K1A3A7A09033389) and by the International Max Planck Research School for Chemistry and Physics of Quantum Materials (IMPRS-CPQM). The authors also acknowledge The Royal Society and The Leverhulme Trust. R. S acknowledges financial support provided by the Ministry of Science and Technology in Taiwan under project number MOST-108-2112-M-001-049-MY2 \& MOST 109-2124-M-002-001 and Sinica funded i-MATE financial Support AS-iMATE-109-13. Access to the Artemis Facility was funded by STFC. The Advanced Light Source is supported by the Director, Office of Science, Office of Basic Energy Sciences, of the U.S. Department of Energy under Contract No. DE-AC02-05CH11231.
\end{acknowledgments}

\newpage

\section{Supplemental Material}

\subsection{1. Experimental details of the photoemission measurements}

Equilibrium angle-resolved photoemission spectroscopy (ARPES) measurements of the ($E_{bin}$, $k_x$, $k_y$, $k_z$)-dependent electronic structure of \VSe2 were carried out at the MAESTRO facility of the Advanced Light Source, using a photon energy range, $h\nu$, from 65 to 160 eV. The energy and angular resolution were set at 20 meV and 0.1$\degree$, respectively.

The time-resolved ARPES (TR-ARPES) experiments were performed at the Artemis facility, Rutherford Appleton Laboratory. Pump and probe beams were provided by a 1 kHz Ti:sapphire laser system, whose output was centered around the fundamental wavelength of 790 nm with a nominal full-width half-maximum (FWHM) pulse duration of 30 fs. A $p$-polarised probe beam was produced via high-harmonic generation (HHG) in a pulsed jet of argon gas, and the 19th harmonic of the HHG spectrum (29.6 eV) was selected with a time-preserving monochromator. The remainder of the beam was used as an $s$-polarised
pump, and its fluence delivered on the sample, 2.2 mJ/cm$^2$, was controlled with a variable neutral density filter. The time and angular resolution of the experiments were set to 40 fs and 0.3$\degree$, respectively. The full width at half maximum (FWHM) of the Gaussian broadening function used for our simulations of the photoemission intensity was around 400~meV in order to achieve a good fit of the data. This energy broadening is a consequence of detector settings chosen to optimise the signal-to-noise ratio. All measurements were performed in ultra-high vacuum conditions. The sample temperature was kept around 70~K, and a clean surface was prepared by cleaving the sample \textit{in situ} prior to beam exposure. 

The equilibrium ARPES measurements were performed in order to determine the three-dimensional (3D) band structure in the hexagonal Brillouin zone (BZ) of 1$T$-VSe$_2$. This was a prerequisite for identifying a suitable $(E_{bin},k_{//})$-cut in the TR-ARPES experiments where the low signal-to-noise ratio and lack of continuous $h\nu$ tunability with the HHG method preclude a detailed investigation of the 3D band structure. 

\begin{figure} [t!]
	\includegraphics[width=0.99\textwidth]{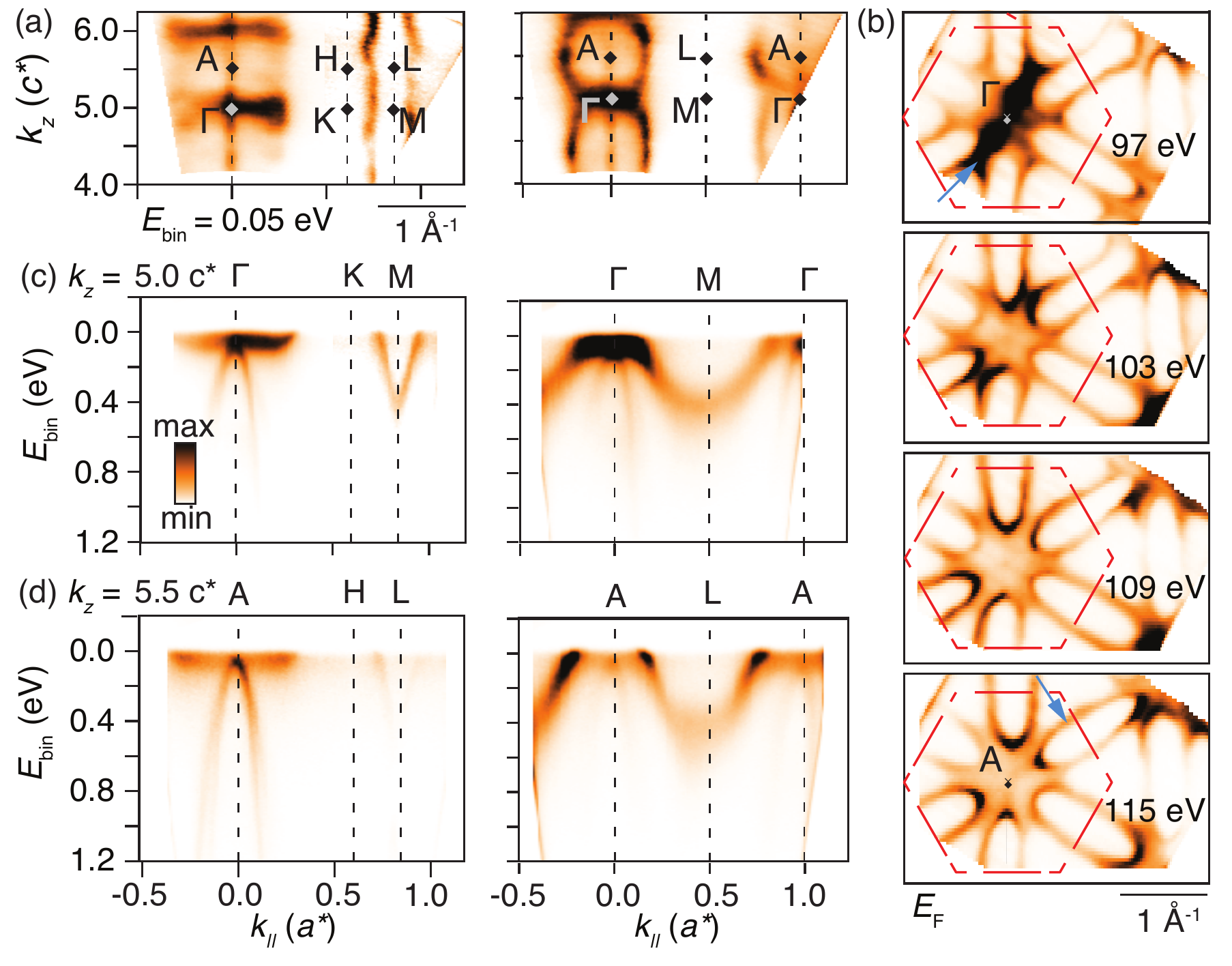}\\
	\caption{(a) ARPES intensity in $(k_{//},k_z)$-cuts around the given high symmetry points for $E_{bin}~=~0.05$~eV. (b) Constant energy contours at the Fermi energy. The red dashed hexagon outlines the BZ.
	Blue arrows highlight central features of the Fermi contour.
	(c) Dispersion along $\mathrm{\Gamma - K - M}$ (left panel) and $\mathrm{\Gamma - M - \Gamma}$ (right panel) obtained for for $k_z = 5.0c^{\ast}$. (d) Analogous data as in (c) for $k_z = 5.5c^{\ast}$.}
	\label{figS1}
\end{figure}

The ARPES $(k_{//},k_z)$-cuts in Fig. \ref{figS1}(a) were measured with the sample oriented in a normal emission geometry with respect to the electron analyzer. In this configuration, the $k_z$ component can be determined from a scan of $h\nu$ using the free-electron final state assumption. At normal emission, these quantities can then be related by the expression $k_z~=~\sqrt{(2m/\hbar^2)[V_0 + (h\nu - E_{bin} - \phi)cos^2\theta]}$ where $m$ is the free-electron mass, $\hbar$ is Planck's constant, $\phi$ is the work function, $\theta$ is the polar emission angle, and $V_0$ is the inner potential. The latter is selected such that the repetition of features that is evident in Figs. \ref{figS1}(a) agrees with the reciprocal lattice parameter $c^{\ast} = 2\pi/c = 1.03$ \AA$^{-1}$. We find that $V_0 = 7.5$~eV gives the best agreement with $c^{\ast}$, in line with previous soft X-ray ARPES experiments \cite{Strocov:2012}.

The evolution of the Fermi surface between $\mathrm{\Gamma}$ and $\mathrm{A}$ is investigated in Fig. \ref{figS1}(b) using fixed $h\nu$ measurements of the $k_{//}$-dependent constant energy contours. Around $\mathrm{\Gamma}$, the intensity concentrates strongly in a bow tie-shaped feature (see blue arrow in Fig. \ref{figS1}(b), top panel), whereas distorted elliptical pockets develop around the $\mathrm{L}$-point (see blue arrow in Fig. \ref{figS1}(b), bottom panel). These features are accompanied by upwards curving bands that cross $E_F$ along $\mathrm{\Gamma-K}$ (Fig. \ref{figS1}(c), left) while downwards curving bands are observed along $\mathrm{\Gamma-M}$ (Fig. \ref{figS1}(c), right). A faint intensity is seen at $E_F$ along $\mathrm{A-H}$ (Fig. \ref{figS1}(d), left) whereas clear $E_F$ crossings occur along $\mathrm{A-L}$ (Fig. \ref{figS1}(d), right). The three-fold symmetry of the material leads to an asymmetric shape of the constant energy contours on two opposite sides of the hexagonal BZ as seen in Fig. \ref{figS1}(b). These features are in overall good agreement with previous ARPES studies  \cite{Terashima:2003, Sato:2004, Strocov:2012}. For the time-resolved measurements, we aligned the $\mathrm{\Gamma-K}$ direction with the analyzer slit. This permits measurements of the excited state signal following the upwards curving bands that cross $E_F$.

\newpage

\subsection{2. Details of the spectral function simulations}

The expression used to model the energy- and momentum-dependent photoemission intensity, ${\cal I}$, is given by \cite{Andreatta:2019}
\begin{equation}
\begin{split}
{\cal I}(E,k_{//}) =
\frac{1}{\pi}\frac{[\mathcal{O}+\mathcal{P}_1E+\mathcal{P}_2E^2+\mathcal{Q}_1k+\mathcal{Q}_2k^2]\times\Sigma^{\prime\prime} }{\left[E-\epsilon(k_{//}) - \Sigma^{\prime} \right]^2 +{\Sigma^{\prime\prime}}^2}
[e^{E/k_BT_e} + 1]^{-1}\\+
\alpha+\beta E(<E_f)+ \gamma E(>E_f).
\end{split}
\label{eq:transphoto}
\end{equation}

The bare band $\epsilon(k_{//})$ is described using the dispersion obtained from density functional theory (DFT). The energy and momentum dependence of the intensity within the bands is modeled using second-order polynomials given by $\sum\limits_{i=0}^{2}\mathcal{P}_i E^i$ and $\sum\limits_{j=0}^{2}\mathcal{Q}_j k^j$. The real and imaginary parts of the self-energy $\Sigma^{\prime}$ and $\Sigma^{\prime\prime}$ are taken as constants within a band. The factor $(e^{E/k_BT_e} + 1)^{-1}$ corresponds to the Fermi-Dirac distribution where $k_B$ is Boltzmann's constant and $T_e$ is the electronic temperature. The background intensity is described by a constant offset, $\alpha$, and allowed to vary linearly with different slopes, $\beta$ and $\gamma$, below and above the Fermi level, respectively. The expression is convoluted with Gaussian functions such that ${\cal I}_{conv}(E,k_{//}) = {\cal I}(E,k_{//})\ast G(\Delta E) \ast G(\Delta k_{//})$ where $\Delta E$ and $\Delta k_{//}$ are the FWHM that correspond to the energy and momentum broadening set by the experimental conditions, respectively.

The above parameters are optimised by fitting a spectrum taken before the optical excitation, where the electronic temperature is equal to the sample temperature. All parameters except  $T_e$ and $\Sigma^{\prime\prime}$ are then held fixed for fits at each time delay. This leads to the fits and time-dependent values of $T_e$ and $\Sigma^{\prime\prime}$ shown in Fig. 3 of the main manuscript.

\newpage

\subsection{3. Three temperature model rate equations}

\begin{figure} [t!]
	\includegraphics[width=0.45\textwidth]{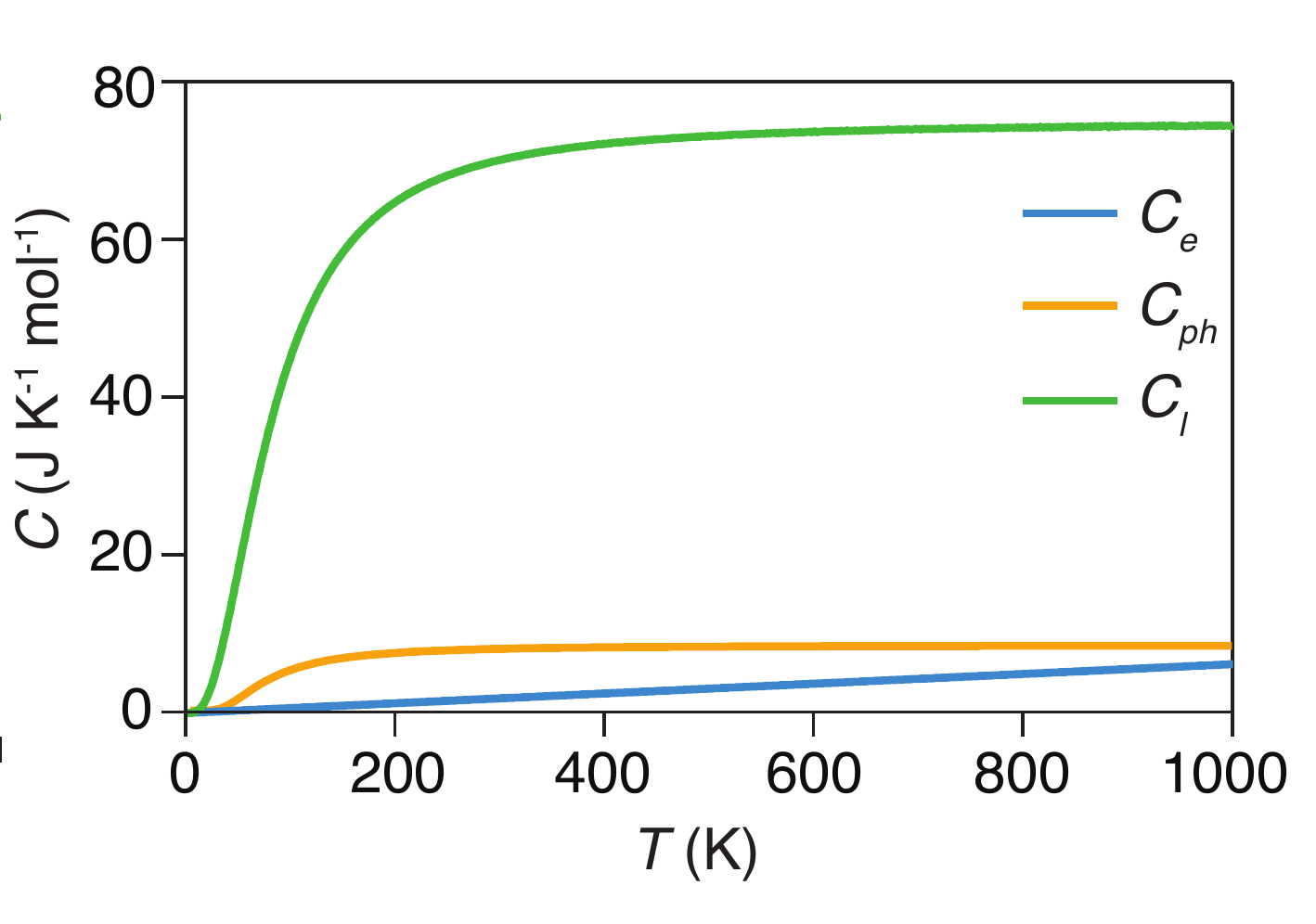}
	\caption{Temperature dependence of heat capacities for electrons (blue), strongly coupled optical phonon with energy 21 meV (orange) and the entire lattice (green curve).}
	\label{figS2}
\end{figure}

Simplifying the energy transfer rate equations as derived by Allen \cite{Allen:1987}, we arrive at the rate equations for the electron and strongly coupled optical phonon temperatures, $T_e$ and $T_{ph}$ \cite{Kampfrath:2005}. We add an anharmonic phonon-phonon coupling term and establish a third rate equation for the lattice temperature, $T_l$ \cite{Perfetti:2007,DalConte:2012,Johannsen:2013}:
\begin{gather}
\frac{dT_e}{dt} = \frac{S(t)}{\beta}-\frac{N_A\,\pi\lambda\, g(E_F)\,[\hbar\omega_{ph}]^3}{\hbar}\frac{n_e-n_{ph}}{C_e},
\label{eq:etemp}\\
\frac{dT_{ph}}{dt} = \frac{C_e}{C_{ph}}\frac{N_A\,\pi\lambda\, g(E_F)[\hbar\omega_{ph}]^3}{\hbar}\frac{n_e-n_{ph}}{C_e}-\frac{T_{ph}-T_l}{\tau_{an}},
\label{eq:phtemp}\\
\frac{dT_l}{dt} = \frac{C_{ph}}{C_l-C_{ph}}\frac{T_{ph}-T_l}{\tau_{an}}.
\label{eq:lattemp}
\end{gather}
Here, $S(t)$ is a Gaussian pulse shape that models the laser field, $\beta$ is a parameter that describes the energy absorbed as heat by the material, $N_A$ is Avogadro's number, $\lambda$ is electron-phonon coupling (EPC) strength, $\tau_{an}$ denotes the anharmonic phonon-phonon decay time, $\omega_{ph}$ is a single Einstein mode phonon frequency, $k_B$ is Boltzmann's constant, $n_e$ ($n_{ph}$) is the Bose-Einstein distribution evaluated at $T_e$ ($T_{ph}$), and $g(E_F) = 2.613$~eV$^{-1}$ is the electronic DOS evaluated at $E_F$, as determined from our DFT calculations. We assume that the temperature dependence of the chemical potential is insignificant in bulk \VSe2 for our electronic temperatures. 
$C_{e}$ is taken as the linear electronic heat capacity, $C_{ph}$ is the heat capacity of a fraction of Einstein modes centered at $\hbar\omega_{ph} = 21$~meV, and $C_l$ is the full lattice heat capacity calculated numerically from the phonon DOS, $F(\omega)$, where we neglect the temperature dependence of the DOS for simplicity, as specified by the following equations:
\begin{gather}
C_e = \frac{\pi^2}{3}\,N_A\,k_B^2\,g(E_F) T_e,
\label{eq:ce}\\
C_{ph} = N_A\, k_B\bigg(\frac{\hbar\omega_{ph}}{k_B T_{ph}}\bigg)^2 \frac{e^{\hbar\omega_{ph}/k_B T_{ph}}}{\big(e^{\hbar\omega_{ph}/k_B T_{ph}}-1\big)^2},
\label{eq:cph}\\
C_l = N_A\,k_B\, \frac{d}{dT_l} \int^{+\infty}_{-\infty} F(\omega)\,\hbar\omega\,n_l(\hbar\omega)\, d\omega.
\end{gather}
Here, $n_l$ is the Bose-Einstein distribution evaluated at $T_l$. Note that the pre-factor in the expression for $C_e(T_e)$ agrees well with the experimental Sommerfeld constant of 7~mJ~K$^{-2}$~mol$^{-1}$~\cite{Yadav:2010}. The temperature dependence of each heat capacity is presented in Fig. \ref{figS2}.

The fit of the three-temperature model rate equations leads to a value of $\beta$ of $5.32~\cdot~10^{-7}$~JK$^{-1}$cm$^{-2}$, which is strongly constrained by the maximum of $T_e$. Furthermore, we find that the EPC constant is given by $\lambda = 1.79 \pm 0.06$ and the fitted anharmonic decay time is  $\tau_{an} = (201 \pm 7)$~fs. The very large value of $\lambda$ is attributed to our choice of limiting the fraction of strongly coupled modes to those at 21~meV in the 3TM, whereas in reality the energy transfer in the early stages of relaxation likely includes a wider distribution of the modes seen in Figs. 4(b)-(c) of the main manuscript.

\newpage

\subsection{4. Details of calculations}
The electronic structure of bulk \VSe2 was calculated within DFT using the Wien2k package \cite{Blaha:2018}. We utilized the linearized augmented plane wave (LAPW) method and the Perdew Burke and Ernzerhof (PBE) exchange-correlation functional \cite{Perdew:1996}. We employed a $12\times12\times6$ $\bm{k}$-point mesh and the energy (charge) convergence was set to be $10^{-5}$~Ry ($10^{-5}$~C).
Experimental lattice parameters have been adopted for band structure calculations \cite{Bayard:1976}. The spin-orbit coupling was neglected in this work.

Phonon calculations were performed by density functional perturbation theory (DFPT) provided in the QUANTUM ESPRESSO code \cite{Giannozzi:2009}. Ultrasoft pseudopotentials were employed to describe interactions between core and valence electrons \cite{Vanderbilt:1990}. The exchange-correlation interaction was treated by the generalised gradient approximation parametrised by the PBE functional. The kinetic energy cutoff for wave functions (charge density) in the plane wave expansion was set to 60 (600) Ry. The atomic structure and the unit cell parameters have been fully relaxed until the Hellmann-Feynman force on each atom was less than $10^{-4}$ Ry/Bohr and the convergence criterion for self-consistent calculations was set to be $10^{-6}$ Ry. Van der Waals interactions between layers were included through semiempirical DFT-D2 implementation of the method of Grimme \cite{Grimme:2006}. The acquired lattice parameters, $a=3.321$ \AA$^{-1}$ and $c=6.256$ \AA$^{-1}$ are in agreement with the experimental values \cite{Bayard:1976}.

For the phonon dispersion calculation we employed a Monkhorst-Pack uniform $24\times24\times12$ $\bm{k}$-point grid and a $8\times8\times4$ $\bm{q}$-point mesh for the dynamical matrices. A narrow soft mode, or Kohn anomaly, appears in the acoustic branches in the $q_z=c^{\ast}/3$ plane (see Fig. 4(a) in the main manuscript) for temperatures below a critical limit. The soft modes reported in previous works \cite{Si:2020,Zhang:2017} in the $q_z=0$ plane are computational artifacts, that we avoided by using a finer $\bm{k}$-point grid (the effect of grid size on the soft modes has been studied by Duong \textit{et al.} \cite{Duong:2015}).
Temperature is incorporated into the calculations using a smearing function with the parameter $\sigma$. 
Checking Fermi-Dirac, Marzari-Vanderbilt-DeVita-Payne (MV) \cite{Marzari:1999} and Gaussian functions, we confirm they all predict the soft mode at the CDW $\bm{q}$-vector, however, with different critical parameters $\sigma_c$. For the phonon calculations in Fig. 4, we have adopted the Fermi-Dirac (FD) smearing with $\sigma=$ 0.0037, 0.0045, 0.0071, 0.01 and 0.015~Ry, corresponding to the temperature values of $T^{\ast}=$ 584, 710, 1121, 1579 and 2368~K, respetively. Note that the theoretical critical temperature, $T^{\ast}_c=$ 561~K, should not be compared with the experimental value, as the approximations to DFT are known to overestimate the absolute value of the critical temperature \cite{Duong:2015}.

\begin{figure} [t!]
	\includegraphics[width=0.98\textwidth]{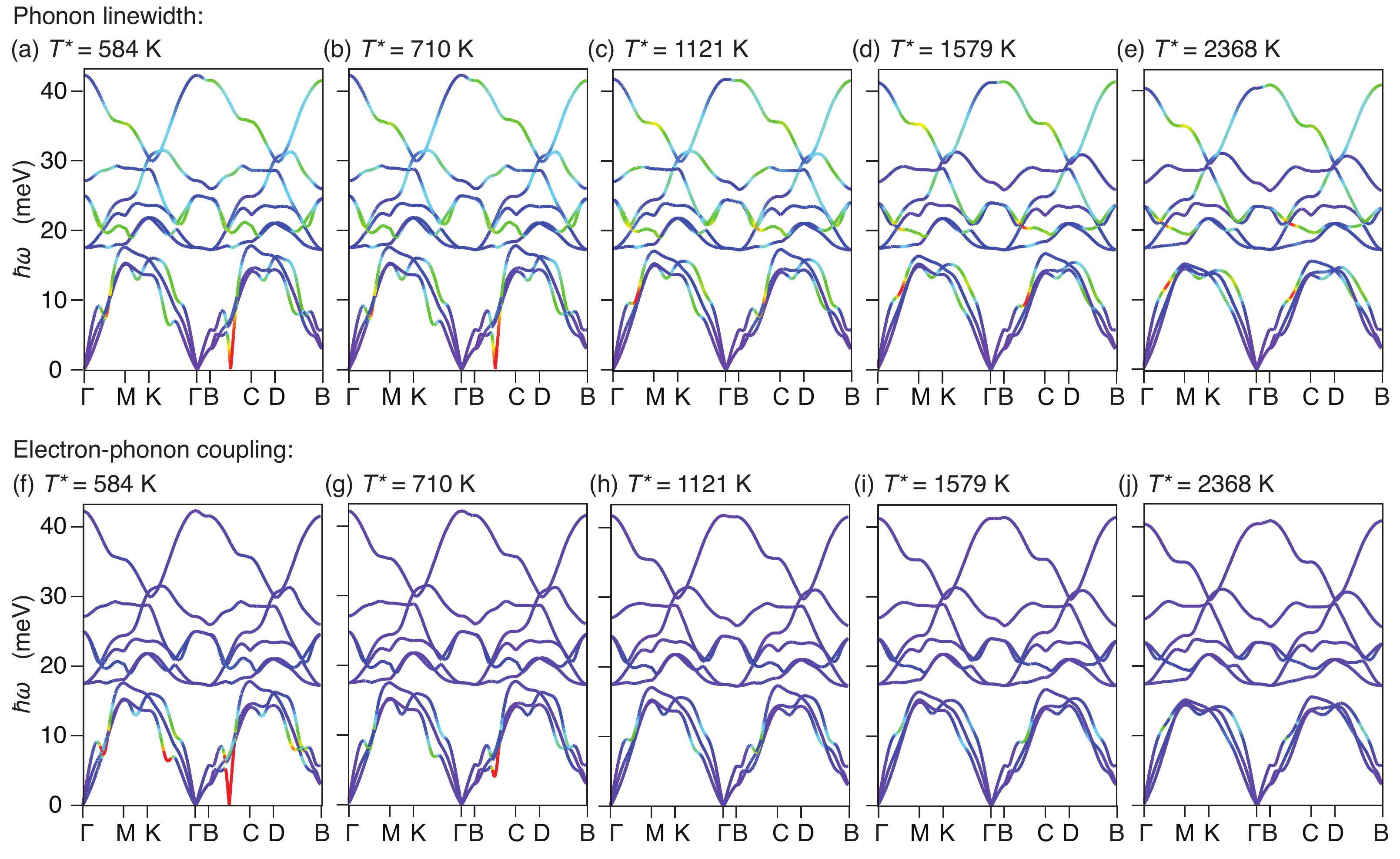}
	\caption{(a)-(e) Phonon linewidth for the system at a range pf Fermi-Dirac smearing temperatures, $T^{\ast}$. (f)-(j) Corresponding electron-phonon coupling.}
	\label{figS3}
\end{figure}

The electron-phonon properties have been obtained with the isotropic approximation to Migdal-Eliashberg theory \cite{migdal1958interaction,eliashberg1960interactions}. For these calculations, a Monkhorst-Pack uniform $16\times16\times8$ $\bm{k}$-point grid was used for the phonon dispersions. The mode resolved EPC constants were calculated on a fine mesh grid containing $32\times32\times16$ $\bm{k}$-points. The convergence of electron-phonon matrix elements was checked with grids up to $48\times48\times24$ in the reciprocal space.
In Fig. \ref{figS3}, we present the phonon dispersions for the aforementioned FD smearing factors, together with associated phonon linewidth, $\gamma$, and EPC strength, $\lambda$, encoded by the color scale. Note the following relationship between these two qualities:
\begin{gather}
\lambda = \frac{\gamma}{\pi g(E_F)\,\omega_{ph}^2}.
\label{eq:linewidth}
\end{gather}

\begin{figure} [t!]
	\includegraphics[width=0.52\textwidth]{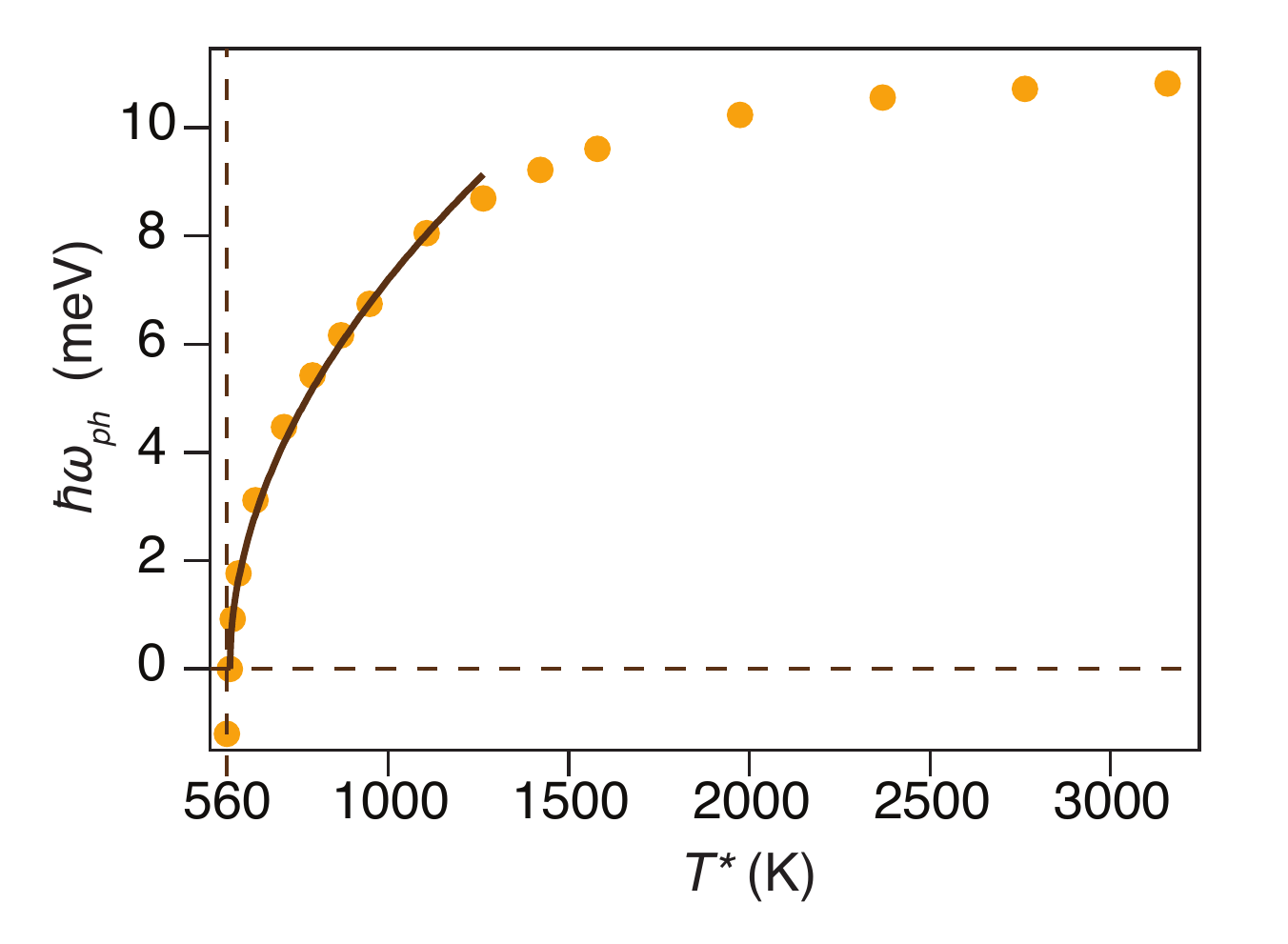}
	\caption{Evolution of the soft phonon mode frequency at the CDW $\bm{q}$-vector as a function of FD smearing temperature (orange markers). Brown solid line indicates a mean-field theory fit (see text).}
	\label{figS4}
\end{figure}

Additionally, we performed single $\bm{q}$-point calculations of the soft phonon frequency at the CDW vector versus the FD smearing temperature, as shown in Fig. \ref{figS4}. We fit the temperature dependence of the frequency according 
to the function $\omega(T^{\ast})=\omega_{0}{T^{\ast}_{red}}^{\delta}$, where $T^{\ast}_{red} =  (T^{\ast} - T^{\ast}_c)/T^{\ast}_c$ is the reduced smearing temperature.
The fit yields $\omega_{0} = 8.8 \pm 0.2$~meV and a critical exponent, $\delta$ = 0.49, which is close to the mean-field theory result for the description of the second-order phase transition \cite{Duong:2015,Diego:2020}, providing evidence that the CDW phase transition does not proceed through an energy barrier \cite{Wall:2012}. Since imaginary frequencies are not physical, we only include the real frequency values in the fit and only account for the frequencies corresponding to temperatures close to the critical point. 

\end{document}